\documentclass[
,draft            
,numberedheadings 
  ]
  {aipproc}
\usepackage{color}

\layoutstyle{8x11single}

\def\be{\begin{equation}}
\def\ee{\end{equation}}
\def\bea{\begin{eqnarray}}
\def\eea{\end{eqnarray}}
\def\d{\mbox{d}}

\def\p{\partial}

\def\pder#1#2{\frac{\partial #1}{\partial #2}}

\let\phi=\varphi
\let\rho=\varrho

\begin{document}

\title{Toroidal LNRF-velocity profiles in thick accretion discs orbiting
       rapidly rotating Kerr black holes} 


\author{Zden\v{e}k Stuchl\'{\i}k}{
  address={Institute of Physics, Faculty of Philosophy and Science, Silesian
  University in Opava, \\
  Bezru\v{c}ovo n\'{a}m. 13, CZ-74601 Opava, Czech Republic}
}

\author{Petr Slan\'{y}}{
}

\author{Gabriel T\"{o}r\"{o}k}{
}

\begin{abstract}
We show that in the equatorial plane of marginally
stable thick discs (with uniformly distributed specific angular momentum
  $\ell(r,\theta)=\mbox{const}$) the
orbital velocity relative to the LNRF has a positive radial gradient in the
vicinity of black holes with $a>0.99979$. The change of sign of the velocity
gradient occurs just above the center of the thick toroidal
discs, in the region where stable circular geodesics of the Kerr spacetime are
allowed. The global character of the phenomenon is given in terms of topology
changes of the von Zeipel surfaces (equivalent to equivelocity surfaces in the
tori with $\ell (r,\theta)=\mbox{const}$). Toroidal von Zeipel surfaces exist
around the circle corresponding to the minimum of the
equatorial LNRF velocity profile, indicating a possibility of development of
some vertical instabilities in those parts of marginally stable tori with
positive gradient of the LNRF velocity. Eventual oscillatory frequencies
connected with the phenomenon are given in a~coordinate-independent form.

\end{abstract}

\maketitle


\section{Introduction}

High frequency (kHz) quasi-periodic oscillations (QPOs) with frequency
ratios 3:2 (and sometimes 3:1) are observed in microquasars
(see, e.g., \cite{Kli:2000:ARASTRA:, %
                  McCli-Rem:2004:CompactX-Sources:}). 
The same frequency ratios of QPOs in mHz are observed in the Galactic Center
black hole Sgr A* \cite{Gen-etal:2003:NATURE:, %
                                Asch:2004:ASTRA:}. 
It is assumed now that the QPOs are related to the parametric or
forced resonance \cite{Lan-Lif:1973:Mech:} of the radial and vertical epicyclic
oscillations or one of the epicyclic and orbital oscillations in accretion
discs
\cite{Now-Leh:1998:TheoryBlackHoleAccretionDisks:, %
      Klu-Abr:2001:ACTPB:, %
      Abr-Klu:2003:GENRG1:}.

The oscillations could be related to both the thin Keplerian discs
\cite{Abr-etal:2003:PUBASJ:, %
      Kat:2001:PUBASJ:}
 or the thick, toroidal accretion discs 
\cite{Rez-etal:2003:MONNR:, %
      Klu-Abr-Lee:2004:X-RayTiming2003:}. 
The parametric resonance of the radial and vertical oscillations in the thin
discs can explain the QPOs with the $\omega_{\theta}/\omega_{r}=3:2$ frequency
ratio observed in all the microquasars and 
can put strong limits on the rotational parameter of their central black holes
related to the limits on their mass
\cite{Tor-Abr-Klu-Stu:2005:ASTRA:}. 

Aschenbach \cite{Asch:2004:ASTRA:} conjectured  
relation between the 3:2 and 3:1 resonance orbits by relating their Keplerian
orbital velocities at $r_{3:2}$ and $r_{3:1}$ to be $\Omega_{\rm
  K}(r_{3:1};a)=3\Omega_{\rm K}(r_{3:2};a)$, fixing thus the rotational
parameter of black holes at the value of $a=a_f=0.99616$. Further, he proposed
that excitation of the oscillations at $r=r_{3:1}$ can be related to two
changes of sign of the radial gradient of the Keplerian orbital velocity as
measured in the locally non-rotating frame (LNRF)
\cite{Bar-Pre-Teu:1972:ASTRJ2:} that occurs in vicinity of $r=r_{3:1}$ for
black holes with $a>0.9953$. While the assumption of frequency
commensurability of Keplerian orbits at 
$r_{3:1}$ and $r_{3:2}$ seems to be rather artificial because distant parts of
the Keplerian disc have to be related, we consider the positive radial
gradient of orbital velocity in LNRF nearby the $r_{3:1}$ orbit around black
holes with $a>0.9953$ to be a physically interesting phenomenon, even if a
direct mechanism relating this to triggering of the excitation of radial (and
vertical) epicyclic oscillations is unknown. 
We also show (Table~\ref{t1})
  that physically relevant, i.e. coordinate-independent, frequencies, given by
maximal positive radial gradient of the LNRF-orbital velocity, take
\emph{locally} values from tens to hundreds of Hz and for stationary observers
at infinity tens of Hz (for typical stellar-mass black holes $M\sim
10\,M_{\odot}$), which likely disables possibility to identify these 
frequencies directly with those of high frequency QPOs
  \cite{Tor-Abr-Klu-Stu:2005:ASTRA:}. 

Because the accretion-disc regime will vary from thin Keplerian
disc to thick toroidal disc with variations of accretion flow, we shall study
here, without addressing details of the mechanism, whether the orbital
velocity in LNRF can have positive gradient also for 
matter orbiting black holes in marginally stable thick discs with uniform
distribution of the specific angular momentum ($\ell(r,\theta)=\mbox{const}$),
leading to a possibility to excite oscillations in the thick-disc accretion
regime.  
Note that the assumption of uniform distribution of the specific angular
momentum can be relevant at least at the inner parts of the 
thick disc and that matter in the disc follows nearly geodesic circular
orbits nearby the centre of the disc and in the vicinity of its inner edge
determined by the cusp of its critical equipotential surface, see
\cite{Abr-Jar-Sik:1978:ASTRA:, %
      Koz-Jar-Abr:1978:ASTRA:}. 

In thick tori, it is necessary to have information about the character
of the Aschenbach's phenomenon also outside the equatorial plane. We shall
obtain such 
information by introducing the notion of von Zeipel radius $\cal R$,
analogical to the radius of gyration $\tilde{\rho}$ introduced for the case of
Kerr spacetimes in the framework of optical geometry by Abramowicz et al.
\cite{Abr-Nur-Wex:1995:CLAQG:}, generalizing in one special way the definition
used for static spacetimes \cite{Abr-Mil-Stu:1993:PHYSR4:}. The von Zeipel
radius is defined in such a way 
that for the marginally stable tori the von Zeipel surfaces, i.e., the
surfaces of constant values of $\cal R$, coincide with surfaces of constant
orbital velocity relative to the LNRF.  

\section{Toroidal marginally stable accretion discs}\label{s2}

The perfect fluid stationary and axisymmetric toroidal discs 
are characterized by 4-velocity field $U^{\mu} = (U^{t},\,0,\,0,\,U^{\phi})$
with $U^{t}=U^{t}(r,\theta),\ U^{\phi}=U^{\phi}(r,\theta)$, and by the
distribution of specific angular momentum $\ell=-U_{\phi}/U_t$.
The angular velocity of orbiting matter, $\Omega=U^{\phi}/U^t$, is then
related to $\ell$ by the formula
\be                                               \label{e4}
     \Omega=-\frac{\ell g_{tt}+g_{t\phi}}{\ell g_{t\phi}+g_{\phi\phi}}.
\ee

The marginally stable toroids are characterized by the uniform distribution of
specific angular momentum $\ell=\ell (r,\,\theta)=\mbox{const}$
and they are fully determined by the spacetime structure through equipotential
surfaces of the potential $W=W(r,\theta)$ defined by the relation
\cite{Abr-Jar-Sik:1978:ASTRA:}
\be                                               \label{e6}
     W-W_{\rm in}=\ln\frac{U_{t}}{(U_{t})_{\rm in}}, \quad
     (U_t)^2=\frac{g_{t\phi}^2 - g_{tt}g_{\phi\phi}}{g_{tt}\ell^2 +
  2g_{t\phi}\ell + g_{\phi\phi}};
\ee
the subscript ``in'' refers to the inner edge of the disc. 

In the Kerr spacetimes with the rotational parameter assumed to be $a>0$, the
relevant metric coefficients in the standard Boyer-Lindquist coordinates read:
\be                                                        \label{e8}
     g_{tt} = -\frac{\Delta - a^2 \sin^2 \theta}{\Sigma}, \quad
     g_{t\phi} = -\frac{2ar\sin^2 \theta}{\Sigma}, \quad
     g_{\phi\phi} = \frac{A\sin^2 \theta}{\Sigma},
\ee
where
\be                                               \label{e11}
     \Delta = r^2-2r+a^2, \quad
     \Sigma = r^2+a^2 \cos^2 \theta, \quad
     A = (r^2+a^2)^2-\Delta a^2 \sin^2 \theta.
\ee
The geometrical units, $c=G=1$, together with putting the mass of the black
hole equal to one, $M=1$, are used to obtain completely dimensionless formulae
hereafter. The relation (\ref{e4}) for the angular velocity of matter orbiting
the black hole acquires the form
\be                                               \label{e13.1}
     \Omega = \Omega(r,\,\theta;\,a,\, \ell) = \frac{(\Delta-a^2 \sin^2
       \theta)\ell + 2ar\sin^2 \theta}{(A-2\ell ar) \sin^2 \theta}
\ee
and the potential W, defined in Eq. (\ref{e6}), has the explicit form
\be                                               \label{e13.2}
     W = W(r,\,\theta;\,a,\,\ell)=
     \frac{1}{2}\ln\frac{\Sigma\Delta\sin^2\theta}{(r^2+a^2-a\ell)^2
     \sin^2\theta - \Delta(\ell-a\sin^2\theta)^2}. 
\ee

\section{The orbital velocity in LNRF}\label{s3}

The locally non-rotating frames are given by the tetrad of 1-forms
\cite{Bar-Pre-Teu:1972:ASTRJ2:}
\be                                                          \label{e14}
     \mathbf{e}^{(t)} = \left(\frac{\Sigma\Delta}{A}\right)^{1/2}
     \mathbf{d}t, \quad                             
     \mathbf{e}^{(r)} = \left(\frac{\Sigma}{\Delta}\right)^{1/2}
     \mathbf{d}r, \quad                              
     \mathbf{e}^{(\theta)} = \Sigma^{1/2}\mathbf{d}\theta, \quad
     \mathbf{e}^{(\phi)} = \left(\frac{A}{\Sigma}\right)^{1/2}\sin\theta
     (\mathbf{d}\phi-\omega\mathbf{d}t),
\ee
where
\be
     \omega = -\frac{g_{t\phi}}{g_{\phi\phi}} = \frac{2ar}{A} \label{e18}
\ee
is the angular velocity of LNRF. 
The azimuthal component of 3-velocity in LNRF reads
\be                                                          \label{e19}
     {\cal V}^{(\phi)}_{\rm LNRF}=\frac{U^{\mu}
       \mathrm{e}^{(\phi)}_{\mu}}{U^{\nu} \mathrm{e}^{(t)}_{\nu}} =
     \frac{A\sin\theta}{\Sigma\sqrt{\Delta}} (\Omega-\omega).   
\ee
Substituting for the angular velocities $\Omega$ and $\omega$ from the
relations (\ref{e13.1}) and (\ref{e18}), respectively, we arrive at the formula
\be                                                          \label{e20}
     {\cal V}^{(\phi)}_{\rm LNRF}=\frac{A(\Delta - a^2 \sin^2 \theta) + 4a^2
      r^2 \sin^2 \theta}{\Sigma\sqrt{\Delta} (A-2a\ell r)\sin\theta}\ell. 
\ee

We focus our investigation to the motion in the equatorial plane,
$\theta=\pi/2$. Formally, this velocity vanishes for $r\to\infty$ and $r\to
r_{+}=1+\sqrt{1-a^2}$, where the event horizon is located, i.e., there must be
a change of its radial gradient for any case of values of the parameters $a$
and $\ell$, contrary to the case of Keplerian orbits characterized by the
Keplerian distributions of the angular velocity and the specific angular
momentum 
\be                                                          \label{e21.1}
     \Omega=\Omega_{\rm K}(r;\,a)\equiv\frac{1}{(r^{3/2}+a)}, \quad
     \ell=\ell_{\rm K}(r;\,a) \equiv
     \frac{r^2-2ar^{1/2}+a^2}{r^{3/2}-2r^{1/2}+a},
\ee
where the azimuthal component of the 3-velocity in LNRF in the equatorial
plane, $\theta=\pi/2$, reads
\be                                                          \label{e22}
     {\cal V}^{(\phi)}_{\rm K}(r;\,a)=\frac{(r^2+a^2)^2 - a^2\Delta -
       2ar(r^{3/2}+a)}{r^2(r^{3/2}+a)\sqrt{\Delta}} 
\ee 
and formally diverges at $r=r_{+}$. 

Of course, for both thick tori and Keplerian discs we must consider the limit
on the disc extension given by the innermost stable orbit. For Keplerian discs
this is the marginally stable geodetical orbit, while for the thick tori this
is an unstable circular geodesic keeped stable by pressure gradients and
located between the marginally bound and the marginally stable geodetical
orbits, with the radius being determined by the specific angular momentum
$\ell=\mbox{const}\in (l_{\rm ms},l_{\rm mb})$ through the equation $\ell=
\ell_{\rm K}(r;\,a)$; $\ell_{\rm ms}$ ($\ell_{\rm mb}$) denotes specific
angular momentum of the circular marginally stable (marginally bound) geodesic.

The radial gradient of the equatorial orbital velocity of thick discs reads
\be                                              \label{e23}
    \pder{{\cal V}^{(\phi)}}{r} =
    \frac{[\Delta+(r-1)r][r(r^2+a^2)-2a(\ell-a)] -
       r(3r^2+a^2)\Delta}{[r(r^2+a^2)-2a(\ell-a)]^2 \sqrt{\Delta}}\ell,
\ee
so that it changes its orientation at radii determined for a given
$\ell\in(l_{\rm ms},l_{\rm mb})$ by the condition
\be                                               \label{e24}
     \ell=\ell_{\rm ex}(r;a) \equiv a +
     \frac{r^2[(r^2+a^2)(r-1)-2r\Delta]}{2a[\Delta+r(r-1)]}.
\ee

Detailed discussion \cite{Stu-Sla-Tor-Abr:2005:PHYSR4:} shows that
two changes of sign of $\p {\cal V}^{(\phi)}/\p r$ can occur for Kerr
black holes with the rotational parameter $a>a_{\rm c(thick)}\doteq
0.99979$. The interval of relevant values of the specific angular momentum
$\ell\in (\ell_{\rm ms}(a),\ell_{\rm ex(max)}(a))$ grows with $a$ growing up
to the critical value of $a_{\rm c(mb)}\doteq 0.99998$. For $a>a_{\rm c(mb)}$,
the interval of relevant values of $\ell\in (\ell_{\rm ms}(a),\ell_{\rm
  mb}(a))$ is narrowing with growing of the rotational parameter up to $a=1$,
which corresponds to a singular case where $\ell_{\rm ms}(a=1)=\ell_{\rm
  mb}(a=1)=2$. Notice that the situation becomes to be singular only in terms
of the specific angular momentum; it is shown
(see \cite{Bar-Pre-Teu:1972:ASTRJ2:})  
that for $a=1$ both the total energy $E$ and the axial angular momentum
$L$ differ at $r_{\rm ms}$ and $r_{\rm mb}$, respectively, but their
combination, $\ell\equiv L/E$, giving the specific angular momentum, coincides
at these radii.  

\section{Von Zeipel surfaces}\label{s4}

\begin{figure}
\centering
\includegraphics[width=.78 \hsize]{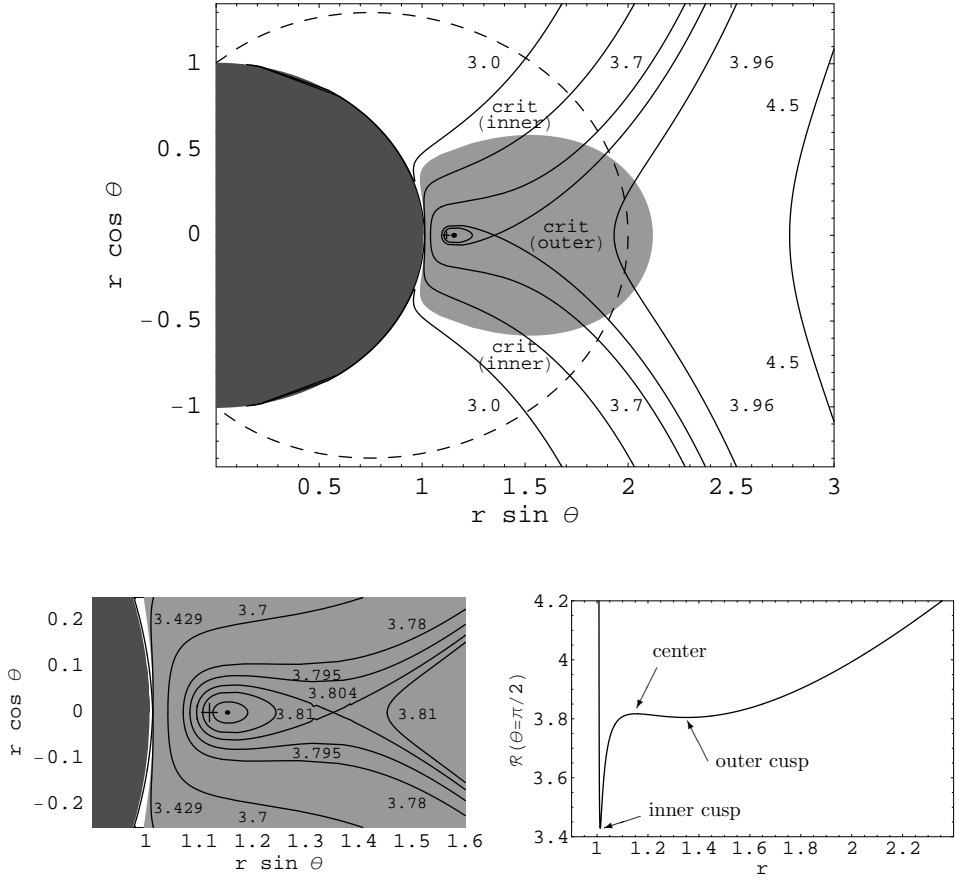}
\caption{Von Zeipel surfaces (meridional sections). For $a<a_{\rm c(bh)}$
  and any $\ell$, only one surface with a cusp in the equatorial plane and no
  closed (toroidal) surfaces exist. The cusp is, however, located outside the
  toroidal equilibrium configurations of perfect fluid. For $a>a_{\rm c(bh)}$
  and $\ell$ appropriately chosen, two surfaces with a cusp, or one surface
  with both the cusps, together with closed 
  (toroidal) surfaces, exist located always inside the ergosphere (dashed
  surface) of a given spacetime. Moreover, if $a>a_{\rm c(thick)}$, both the
  outer  
  cusp and the central ring of closed surfaces are located inside the toroidal
  equilibrium configurations corresponding to marginally stable thick discs
  (light-gray region; its shape is determined by the critical self-crossing
  {\em equipotential surface} of the potential $W$ given by
  (\ref{e13.2})). The cross ($+$) denotes the center of the torus. Dark region
  corresponds to the black hole. Figures illustrating all possible
  configurations of the von Zeipel surfaces are presented in
  \cite{Stu-Sla-Tor-Abr:2005:PHYSR4:}. Here we present  
  the figure plotted for the parameters $a=0.99998,\ \ell=2.0065$. Critical
  value of the von Zeipel radius corresponding to the inner and the outer
  self-crossing surface is ${\cal R}_{\rm c(in)}\doteq 3.429$ and ${\cal
  R}_{\rm c(out)}\doteq 3.804$, respectively, the central ring of toroidal
  surfaces corresponds to the value ${\cal R}_{\rm center}\doteq
  3.817$. Interesting region containing both the cusps and the toroidal
  surfaces is plotted in detail at the left lower figure. Right lower figure
  shows the behaviour of the von Zeipel radius in the equatorial plane.} 
\label{f1}
\end{figure}

It is useful to find global characteristics of the phenomenon that is
manifested in the equatorial plane as the existence of a small region
with positive gradient of the LNRF velocity. A physically reasonable way of
defining a global quantity characterizing rotating fluid configurations in
terms of the LNRF orbital velocity is to introduce, so-called, von Zeipel
radius defined by the relation  
\be                                                      \label{e36}
     {\cal R}\equiv\frac{\ell}{{\cal V}^{(\phi)}_{\rm LNRF}} = 
     (1-\omega\ell)\tilde{\rho}
\ee
which generalizes in another way as compared with
\cite{Abr-Nur-Wex:1995:CLAQG:} the Schwarzschildian definition of the
gyration radius $\tilde\rho$. 
(For more details see \cite{Stu-Sla-Tor-Abr:2005:PHYSR4:}.)

In the case of marginally stable tori with $\ell(r,\theta)=\mbox{const}$, the
von Zeipel surfaces, i.e., the surfaces of ${\cal R}(r,\theta;a,\ell) =
\mbox{const}$, coincide with the equivelocity surfaces ${\cal V}^{(\phi)}_{\rm
  LNRF}(r,\,\theta;\,a,\,\ell)=\mbox{const}$. For the tori in the Kerr
spacetimes, there is
\be                                                      \label{e38}
     {\cal R}(r,\,\theta;\,a,\,\ell) = \frac{\Sigma\sqrt{\Delta}(A-2a\ell
       r)\sin\theta}{A(\Delta-a^2\sin^2\theta)+4a^2r^2\sin^2\theta}.
\ee
Topology of the von Zeipel surfaces can be directly determined by the
behaviour of the von Zeipel radius (\ref{e38}) in the equatorial plane
\be                                                      \label{e39}
     {\cal R}(r,\,\theta=\pi/2;\,a,\,\ell) =
     \frac{r(r^2+a^2)-2a(\ell-a)}{r\sqrt{\Delta}}.
\ee
The local minima of the function (\ref{e39}) determine loci of the cusps of
the von Zeipel surfaces, while its local maximum (if it exists) determines a
circle around which closed toroidally shaped von Zeipel surfaces are
concentrated (see Fig.~\ref{f1}). Notice that the minima (maximum) of
${\cal R}(r,\,\theta=\pi/2;\,a,\,\ell)$ correspond(s) to the maxima (minimum)
of ${\cal V}^{(\phi)}_{\rm LNRF}(r,\,\theta=\pi/2;\,a,\,\ell)$, therefore, the
inner cusp is always physically irrelevant being located outside of the
toroidal configuration of perfect fluid. Behaviour of the von Zeipel surfaces
nearby the center and the inner edge of the thick tori orbiting Kerr black 
holes with $a>a_{\rm c(thick)}\doteq 0.99979$, i.e., the existence of the von
Zeipel surface with an outer cusp or the surfaces with toroidal topology,
suggests 
possibility of strong instabilities in both the vertical and radial direction
and a tendency for development of some vortices crossing the equatorial
plane. We plan studies of these expected phenomena in future.

\begin{figure}
\centering
\includegraphics[width=.8 \hsize]{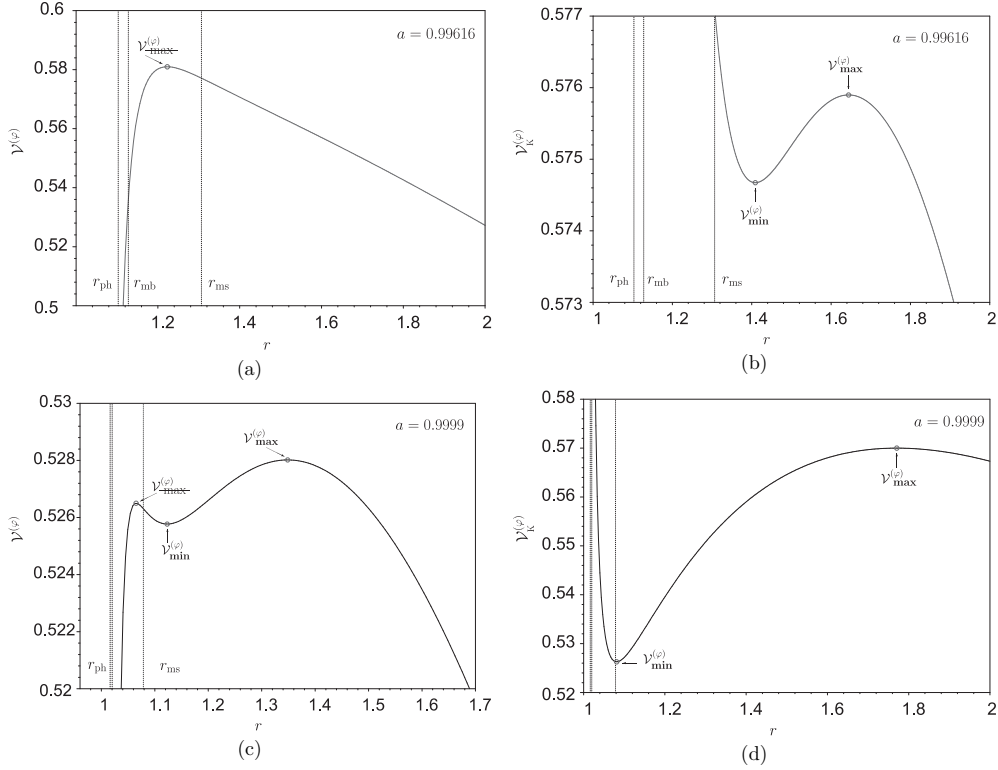}
\caption{Profiles of the equatorial orbital velocity of marginally stable tori
  in LNRF in terms of the radial Boyer-Lindquist coordinate (left column). For
  comparison, 
  the profiles of the orbital velocity of Keplerian discs in Kerr spacetimes
  with the same rotational parameter $a$ are added (right column). For thick
  discs, values of 
  $\ell={\rm const}$ are appropriately chosen; commonly, $\ell=\ell_{\rm ms}$
  is used giving the maximal value of the velocity difference in between the
  local extrema, and representing the limiting case of marginally stable thick
  discs.}
\label{f2}
\end{figure}

\section{Discussion and conclusions}\label{concl}

It is useful to discuss both the qualitative and quantitative aspects of the
phenomenon of the positive gradient of LNRF orbital velocity.
In the Kerr spacetimes with $a>a_{\rm c(thick)}$, changes of sign of the
gradient of ${\cal V}^{(\phi)}(r;\,a)$ must occur closely above the center of
relevant toroidal discs, at radii corresponding to stable circular geodesics
of the spacetime, where the radial and vertical epicyclic frequencies are also
well defined. 

\begin{figure}
\begin{minipage}{.45 \hsize}
\centering
\includegraphics[width=.98 \hsize]{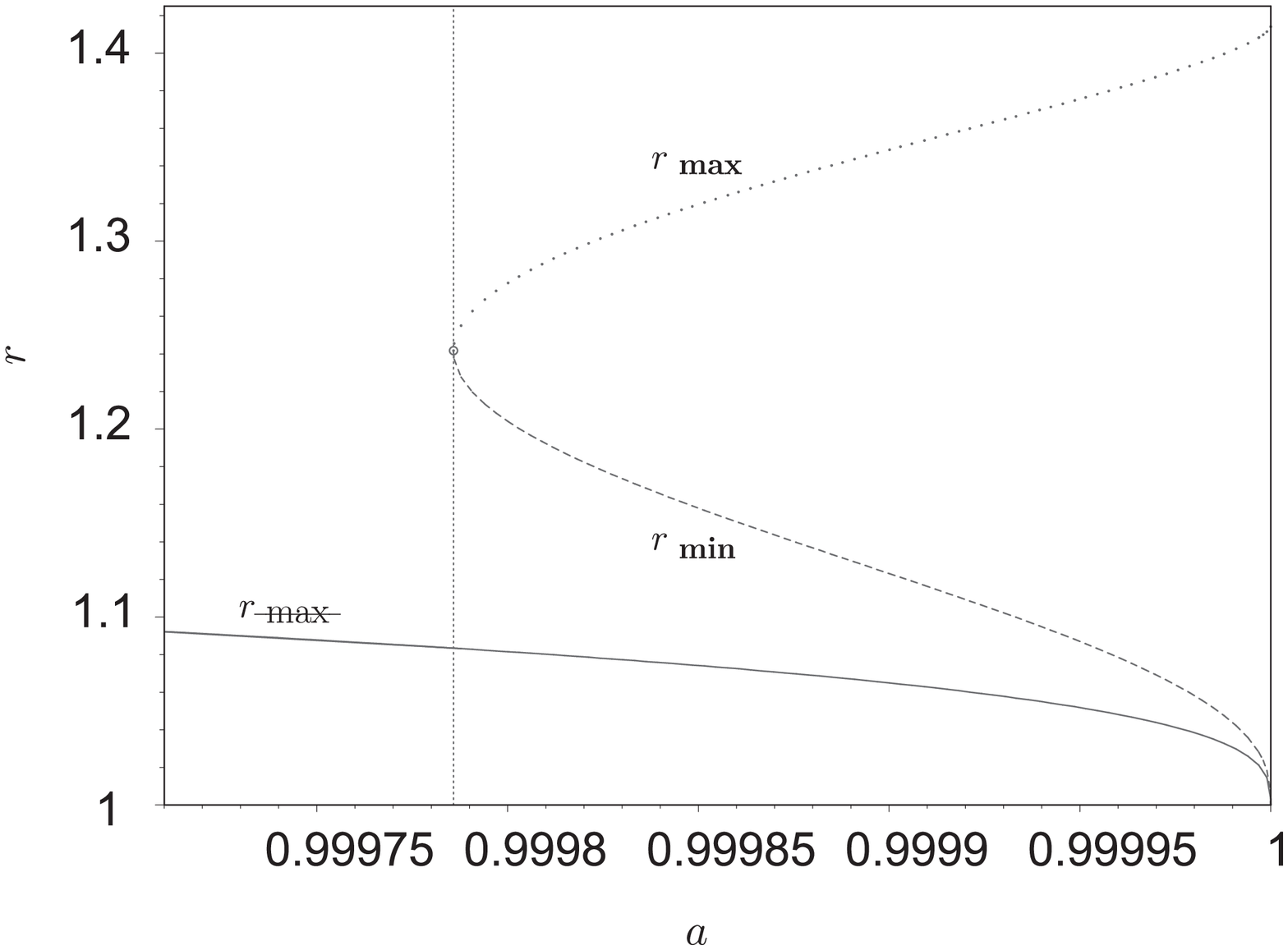}
\par\small (a)\enspace
\end{minipage}\hfill %
\begin{minipage}{.45 \hsize}
\centering
\includegraphics[width=.98 \hsize]{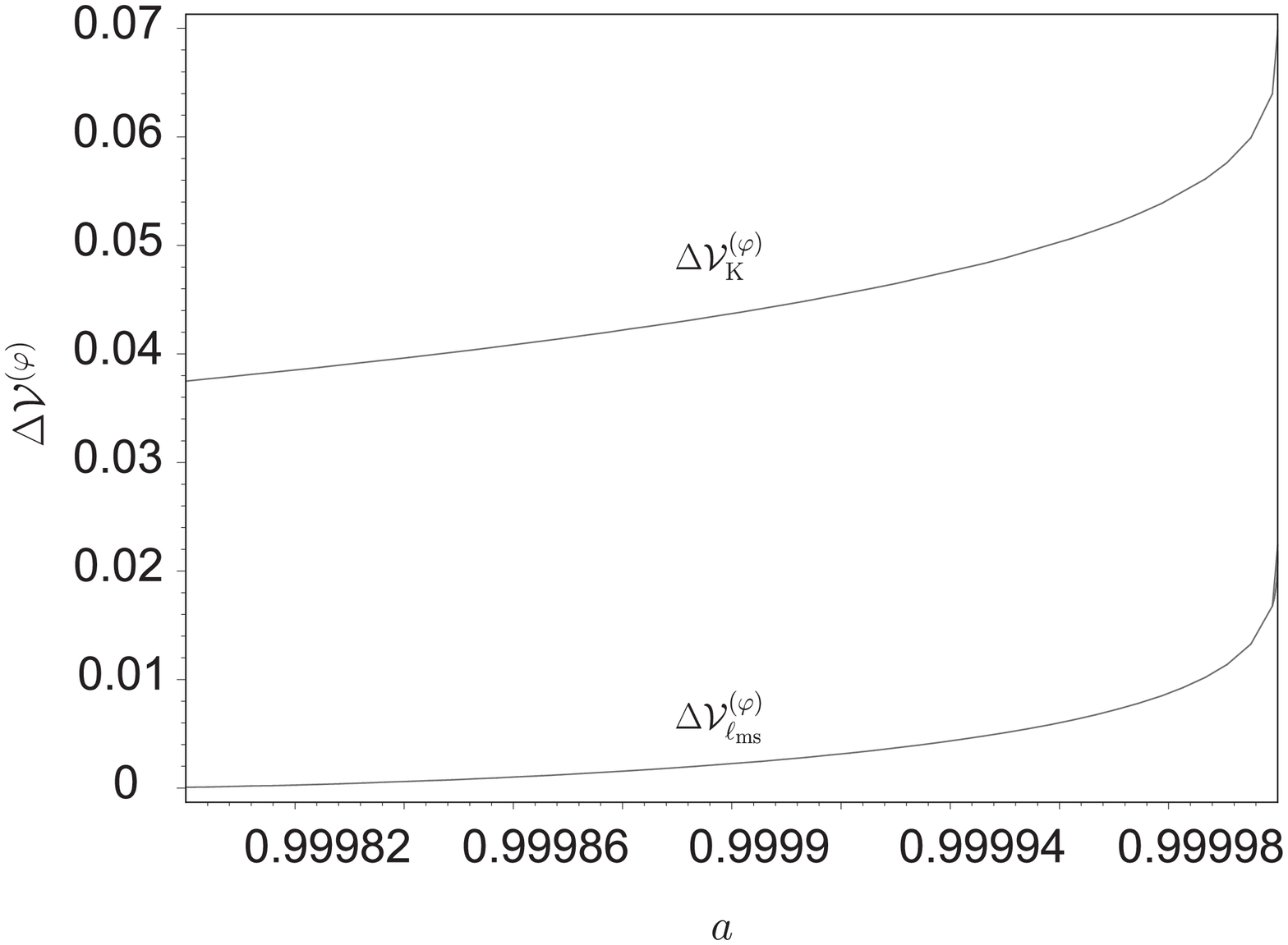}
\par\small (b)\enspace
\end{minipage}
\caption{(a) Positions of local extrema of ${\cal V}^{(\phi)}_{\rm
    LNRF}$ (in B-L coordinates) for the marginally stable discs with
    $\ell=\ell_{\rm ms}$ in dependence on the rotational parameter $a$ of the
    black hole. (b) 
   Velocity difference $\Delta {\cal V}^{(\phi)}={\cal V}^{(\phi)}_{\rm
    max}-{\cal V}^{(\phi)}_{\rm min}$ as a function of the rotational
    parameter $a$ of the 
    black hole for both the thin (Keplerian) disc and the marginally stable
    (non-Keplerian) disc with $\ell=\ell_{\rm ms}$.}
\label{f3}
\end{figure}

In two interesting cases, behaviour of ${\cal V}^{(\phi)}(r;\,a,\,\ell)$ is
illustrated in Fig.~\ref{f2}; for
comparison, profiles of the Keplerian velocity ${\cal V}^{(\phi)}_{\rm
  K}(r;a)$ are included. With $a$ growing in the region of $a\in (a_{\rm
  c(thick)},1)$, the difference $\Delta {\cal V}^{(\phi)}\equiv {\cal
  V}^{(\phi)}_{\rm max}-{\cal V}^{(\phi)}_{\rm min}$ grows as well as the
difference of radii, $\Delta r \equiv r_{\rm max}-r_{\rm min}$, where the
local extrema of ${\cal V}^{(\phi)}(r;a,\ell)$ occur, see Fig.~\ref{f3}.

In terms of the redefined rotational parameter $(1-a)$, the changes of sign of
gradient of the LNRF orbital velocity of marginally stable thick discs occur
for discs orbiting Kerr black holes with $(1-a)<1-a_{\rm c(thick)}\doteq
2.1\times 10^{-4}$ which is more than one order lower than
the value $1-a_{\rm c(thin)}\doteq 4.7\times 10^{-3}$ found by Aschenbach
\cite{Asch:2004:ASTRA:} for
the changes of sign of the gradient of the orbital velocity in Keplerian, thin
discs. Moreover, in the thick discs, the velocity difference, $\Delta {\cal
  V}^{(\phi)}={\cal V}^{(\phi)}_{\rm max}-{\cal V}^{(\phi)}_{\rm min}$, is
smaller but comparable with those in the thin discs (see Fig.~\ref{f3}). In
fact, we can see that for $a \to 1$, the velocity difference in 
the thick discs $\Delta {\cal V}^{(\phi)}_{\rm (thick)}\approx 0.02$, while
for the Keplerian discs it goes even up to $\Delta {\cal V}^{(\phi)}_{\rm
  (thin)}\approx 0.07$. These are really huge velocity differences, being 
expressed in units of $c$.

\begin{figure}
\centering
\includegraphics[width=.8 \hsize]{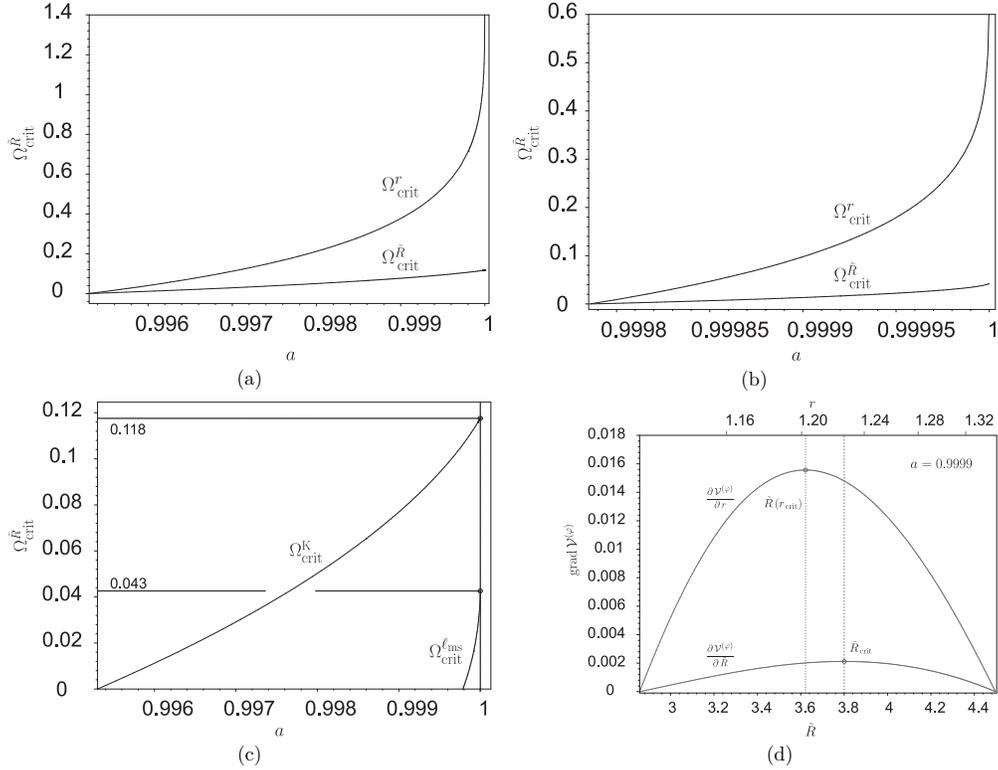}
\caption{Critical ``oscillatory''frequency for excitation of epicyclic
  oscillations, introduced by Aschenbach \cite{Asch:2004:ASTRA:}, as a
  function of the 
  rotational parameter of the black hole in terms of both the B--L coordinate
  radius ($\Omega^{r}_{\rm 
  crit}$) and the proper radial distance ($\Omega^{\tilde{R}}_{\rm
  crit}$). (a) Keplerian discs. (b) Marginally stable (non-Keplerian) discs
  with constant specific angular momentum $\ell=\ell_{\rm ms}$. (c) Comparison
  of critical frequencies for Keplerian $\Omega^{\rm K}_{\rm crit}$ and
  non-Keplerian $\Omega^{\ell_{\rm ms}}_{\rm crit}$ discs in terms of the
  proper radius. (d) Positive parts of the ``coordinate'' and ``proper'' radial
  gradient $\p {\cal V}^{(\phi)}/\p r$ and $\p {\cal V}^{(\phi)}/\p
  \tilde{R}$ for a given value of the rotational parameter $a$.}
\label{f4}
\end{figure}

Following Aschenbach \cite{Asch:2004:ASTRA:}, we can define the typical
frequency of 
the mechanism for excitation of oscillations by the maximum slope of the
positive gradient of $\p {\cal V}^{(\phi)}/\p r$ in the region of the changes
of its sign by the relation
\be                                                        \label{e40}      
     \Omega^{r}_{\rm crit}=2\pi\frac{\p {\cal V}^{(\phi)}}{\p r}|_{\rm max}. 
\ee
The ``oscillatory'' frequency has to be determined numerically. We have done
it for both Keplerian discs and the marginally stable discs with
$\ell=\ell_{\rm ms}=\mbox{const}$, see Fig. \ref{f4} and Table \ref{t1}. It is
interesting that in Keplerian discs with $a\sim 0.996$, there is
$\Omega^{r}_{\rm crit} \sim \Omega_{\rm R} \sim \Omega_{\rm V}/3$, i.e.,
$\Omega^{r}_{\rm crit}$ can be related to the resonant phenomena at the radius
where the epicyclic frequencies are in the 3:1 ratio \cite{Asch:2004:ASTRA:}.
However, it is more correct to consider $\p {\cal V}^{(\phi)}/\p \tilde{R}$
where $\tilde{R}$
is the physically relevant (coordinate-independent) proper radial distance, as
it is more convenient for estimation of physically realistic characteristic
frequencies related to local physics in the disc. Then the critical frequency
for possible excitation of oscillations is given by the relation
\be                                                        \label{e42}
     \Omega^{\tilde{R}}_{\rm crit} = 2\pi\frac{\p {\cal V}^{(\phi)}}{\p
       \tilde{R}}|_{\rm max}, \quad
     \d \tilde{R} = \sqrt{g_{rr}}\d r = \sqrt{\frac{\Sigma}{\Delta}}\d r.
\ee 
Of course, such a locally defined ``oscillatory'' frequency, confined to the
orbiting LNRF-observers, should be further related to distant observers by the
relation (taken at B--L coordinate $r$ corresponding to $(\p {\cal
  V}^{(\phi)}/\p r)_{\rm max}$)
\be                                                           \label{e43}
       \Omega^{\tilde{R}}_{\infty} = \sqrt{-(g_{tt}+2\omega
       g_{t\phi}+\omega^2 g_{\phi\phi})} \Omega^{\tilde{R}}_{\rm crit}.
\ee
Similarly, an analogical relation can be written also for the
critical frequency  $\Omega^{r}_{\rm crit}$, giving the circular frequency
$\Omega^{r}_{\infty}$. 
Because the velocity gradient related to the proper distance is suppressed in
comparison with those related to the coordinate distance, there
is $\Omega^{\tilde{R}}_{\rm crit} < 
\Omega^{r}_{\rm crit}$. The situation is illustrated in
Fig.~\ref{f4}. Moreover, Fig.~\ref{f4}d shows mutual behaviour of the
``coordinate'' and ``proper'' radial gradient $\p {\cal V}^{(\phi)}/\p r$ and
$\p {\cal V}^{(\phi)}/\p \tilde{R}$ in region between the local minimum and
the outer local maximum of the orbital velocity ${\cal V}^{(\phi)}$ for an 
appropriately choosen value of the rotational parameter $a$. 
Characteristic frequencies $f=\Omega/2\pi$, where $\Omega$ corresponds to the
particular circular frequencies $\Omega^{r}_{\rm crit},\ \Omega^{r}_{\infty},\
\Omega^{\tilde{R}}_{\rm crit},\ \Omega^{\tilde{R}}_{\infty}$, respectively, are
given in Table \ref{t1}.

We can conclude that in constant specific angular momentum tori, the
effect discovered by Aschenbach is elucidated by topology changes of the von
Zeipel surfaces. In addition to 
one self-crossing von Zeipel surface existing for all values of the rotational
parameter $a$, for $a>a_{\rm c(thick)}$ the second self-crossing surface
together with toroidal surfaces occur. Toroidal von Zeipel surfaces exist
under the newly developing cusp, being centered around the circle
corresponding to the minimum of the equatorial LNRF velocity profile.
Further, the behaviour of von Zeipel surfaces in
marginally stable tori orbiting Kerr black holes with $a>a_{\rm c(thick)}$
strongly suggests a possibility of development of both the vertical and
vortical instabilities because of the existence of the critical surface with a
cusp, located above the center of the torus and the toroidal von Zeipel
surfaces located under the cusp. The effect of ``velocity gradient sign
changes'' can be very 
important as a trigger instability mechanism for oscillations observed in
QPOs. Of course, further studies directed both to the theoretically well
founded, detailed physical mechanisms for triggering of oscillations in the
equilibrium tori with general specific angular momentum distribution, and the
link to observations, are necessary and planned for the future.

\begin{table}
\centering
\caption{Characteristic frequencies in units
  $(M/M_{\odot})^{-1}$\,Hz ($M/M_{\odot}$ is the mass of the Kerr black hole 
  in units of mass of the Sun.) related to the circular frequencies
  $\Omega=2\pi f$ defined in the text are given for appropriate values of the
  black hole spin. Maximal values of the frequencies related to the stationary
  observer at infinity are bold-faced.}
\begin{tabular}{rcccccccccc}
\hline
& & \multicolumn{4}{c}{Keplerian discs} & & \multicolumn{4}{c}{Fluid tori} \\
\multicolumn{1}{c}{$1-a$} & & $f^{r}$ & $f^{r}_{\infty}$ & $f^{\tilde{R}}$ &
$f^{\tilde{R}}_{\infty}$ & & $f^{r}$ & $f^{r}_{\infty}$ & $f^{\tilde{R}}$ &
$f^{\tilde{R}}_{\infty}$ \\
\hline
$4.5\times 10^{-3}$ & & 356 & 86 & 121 & 29 & & & & \\
$4\times 10^{-3}$ & & 1303 & 303 & 432 & 102 & & & & \\
$3\times 10^{-3}$ & & 3617 & 767 & 1130 & 248 & & \multicolumn{4}{c}{\em
  undefined} \\
$1\times 10^{-3}$ & & 12179 & 1849 & 3061 & 536 & & & & \\
$5\times 10^{-4}$ & & 17132 & 2126 & 3789 & 592 & & & & \\
$2\times 10^{-4}$ & & 22982 & {\bf 2203} & 4352 & {\bf 607} & & 296 & 34 & 57 & 7 \\
$1\times 10^{-4}$ & & 26857 & 2126 & 4579 & 603 & & 3160 & 315 & 555 & 61 \\
$1\times 10^{-5}$ & & 36593 & 1565 & 4816 & 590 & & 10940 & {\bf 657} & 1447 & 135 \\
$1\times 10^{-6}$ & & 42556 & 1001 & 4841 & 588 & & 16271 & 589 & 1718 & 147 \\
$1\times 10^{-9}$ & & 49250 & 201 & 4844 & 588 & & 23277 & 185 & 1807 & {\bf 150} \\
\hline
\end{tabular}
\label{t1}
\end{table}

Finally, we would like to call attention to the fact that signs' changes of the
radial gradient of orbital velocity in LNRF occur nearby the $r=r_{3:1}$
orbit, while in the vicinity of the $r=r_{3:2}$ orbit, $\p {\cal
  V}^{(\phi)}_{\rm LNRF}/\p 
r<0$ for all values of $a$ for both the Keplerian discs and the marginally
stable toroidal discs with all allowed values of $\ell$. Clearly, the
parametric resonance, which is the strongest one for ratios of the epicyclic
frequencies $\Omega_{\rm V}/\Omega_{\rm 
  R}=3/2$ works at the $r=r_{3:2}$ orbit, while its effect is much smaller at
the radius $r=r_{3:1}$ with $\Omega_{\rm V}/\Omega_{\rm R}=3/1$
\cite{Abr-etal:2003:PUBASJ:}. Therefore, 
the forced resonance, triggered by the changes of $\p {\cal V}^{(\phi)}_{\rm
  LNRF}/\p r$, will be important for the 3:1 resonance. Notice that
the forced resonance at $r=r_{3:1}$ can generally result in observed QPOs
frequencies with 3:2 ratio due to the beat frequencies allowed for the forced
resonance \cite{Abr-etal:2004:RAGtime4and5:}; however it seems to be
irrelevant in the case of microquasars, as all observed frequencies lead
to the values of the rotational parameter $a<a_{\rm c(thick)}$ as shown in
\cite{Tor-Abr-Klu-Stu:2005:ASTRA:}.


\begin{theacknowledgments}
This work was supported by the Czech grant MSM~4781305903.
\end{theacknowledgments}



\bibliographystyle{aipproc}   


\IfFileExists{\jobname.bbl}{}
 {\typeout{}
  \typeout{******************************************}
  \typeout{** Please run "bibtex \jobname" to optain}
  \typeout{** the bibliography and then re-run LaTeX}
  \typeout{** twice to fix the references!}
  \typeout{******************************************}
  \typeout{}
 }

\end{document}